\documentclass[12pt]{article}

\usepackage{graphicx}
\usepackage{amsfonts}
\def\+{{+\!\!\!+}}

\def\d{\partial}

\def\pmb#1{\setbox0=\hbox{#1}% 
\kern.0em\copy0\kern-\wd0 
\kern-.04em\copy0\kern-\wd0 
\kern.08em\copy0\kern-\wd0 
\kern-.04em\raise.0433em\box0 }         %poor man's bold macro (TexBook) 
\newcommand{\nc}{\newcommand} 
\nc{\beq}{\begin{equation}} 
\nc{\eeq}[1]{\label{#1}\end{equation}} 
\nc{\ber}{\begin{eqnarray}} 
\nc{\eer}[1]{\label{#1}\end{eqnarray}} 
\nc{\pek}[1]{\cite{#1}} 
\nc{\enr}[1]{(\ref{#1})} 
\nc{\kal}[1]{{\cal{#1}}} 
\nc{\dott}{\;\cdot\;} 
\def\0 {\nonumber}

\begin{document} 
\setcounter{page}{0}
\newcommand{\inv}[1]{{#1}^{-1}} %inverse 
\renewcommand{\theequation}{\thesection.\arabic{equation}} 
\newcommand{\be}{\begin{equation}} 
\newcommand{\ee}{\end{equation}} 
\newcommand{\bea}{\begin{eqnarray}} 
\newcommand{\eea}{\end{eqnarray}} 
\newcommand{\re}[1]{(\ref{#1})} 
\newcommand{\qv}{\quad ,} 
\newcommand{\qp}{\quad .} 
\begin{titlepage} 
%\title{} 
\begin{center} 

\hfill SISSA 74/2006/FM\\     
\hfill UUITP-16/06\\                        
\hfill   hep-th/0611327\\

\vskip .3in \noindent 

%\vskip .1in 

{\Large \bf{Computing Amplitudes in topological M-theory}} \\
%\
%{\Large \textit{alias The Frankenstein Membrane}} \\

\vskip .2in 

{\bf Giulio Bonelli$^{a}$}, {\bf Alessandro Tanzini$^a$} 
and {\bf Maxim Zabzine$^{b}$,}

\vskip .05in 
$^a${\em\small International School of Advanced Studies (SISSA) and INFN, Sezione di Trieste\\
 via Beirut 2-4, 34014 Trieste, Italy} 
\vskip .05in 
$^b${\em  Department of Theoretical Physics, \\
Uppsala University,
Box 803, SE-751 08 Uppsala, Sweden }
\vskip .5in
\end{center} 
\begin{center} {\bf ABSTRACT }  
\end{center} 
\begin{quotation}\noindent  
We define a topological quantum membrane theory on a seven dimensional 
 manifold of $G_2$ holonomy. 
We describe in detail the path integral evaluation 
for membrane geometries given by circle bundles over Riemann surfaces.
We show that when the target space is $CY_3\times S^1$ 
quantum amplitudes of 
non-local observables of
membranes wrapping the circle reduce to the  A-model amplitudes.
 In particular for genus zero we show that our model computes the Gopakumar-Vafa invariants.
Moreover, for membranes wrapping calibrated homology spheres in the $CY_3$, we find
that the amplitudes of our model are related to Joyce invariants. 
\end{quotation} 
\vfill 
\eject 

%\addtocontents{toc}
%\tableofcontents

\end{titlepage}

\section{Introduction}

Topological strings on Calabi-Yau (CY) manifolds describe a certain sector of the superstrings. 
In particular, various BPS quantities in superstrings can be computed using the topological version.
There are two different topological string models on CY, A- and B-models, which can be obtained 
from the physical model through a topological twist  \cite{Witten:1991zz}. 
Different superstring theories are unified in the context of M-theory, which is expected to 
correspond to a supermembrane theory, yet unknown. 
Thus, in analogy with topological strings, one may expect that 
there exists the topological version of M-theory which should capture the BPS sector 
     of M-theory. In particular it is natural to expect that the topological version of M-theory should 
      count  membrane instantons \cite{Becker:1995kb} which induce nonzero corrections to the superpotential.
   The notion of topological M-theory\footnote{Indeed, A.Losev \cite{losev} has advocated the idea of M-theory in the context of the topological string even before these works.} has been proposed in \cite{Dijkgraaf:2004te} (also for earlier proposal see
 \cite{Gerasimov:2004yx}). The idea is that the topological M-theory should provide the unifying description of A- and B-models on $CY_3 \times S^1$. In general the theory should be defined over 
  seven dimensional  $G_2$ manifolds.  In \cite{Dijkgraaf:2004te} the analysis has been done at the classical level of effective actions.  
 Different arguments in favor of topological M-theory have been proposed \cite{Grassi:2004xr, Baulieu:2004pv, Nekrasov:2004vv}.  
 
 Our goal is to propose a microscopic description of topological M-theory in term of topological membrane theory.  This work is the continuation of our previous paper \cite{Bonelli:2005rw}
   (also see the earlier work \cite{Bonelli:2005ti}).  
Different attempts with a similar aim can be found in \cite{Anguelova:2005cv,Bao:2005yx,
Ikeda:2006pd}.

 In our attempt we closely follow the analogy with A-model which is a topological sigma
   model \cite{Witten:1988xj}
  coupled to two dimensional gravity. For the computation in the zero genus sector it is enough to 
   consider the topological sigma model since there are no gravitational moduli for $S^2$. The topological 
    sigma model can be thought of as the BRST quantization \cite{Baulieu:1989rs} 
     of the following topological action
      \beq
       S_{top} = \int\limits_{\Sigma_2} X^*(K),
      \eeq{topAmodel}
        where $K$ is the K\"ahler form on the CY target.  
         We would like to construct a membrane analog of the topological sigma model. We propose 
          that on $G_2$ 
       manifolds\footnote{In this paper we consider the topological membrane theory defined on 
       a seven manifold of  $G_2$ holonomy.  However, the $G_2$ holonomy condition can be relaxed 
       to a closed $G_2$-structure, see further details in \cite{Bonelli:2005rw}.} 
        this model arises as the BRST quantization of the topological action
       \beq
     S_{top} = \int\limits_{\Sigma_3} X^*(\Phi),
     \eeq{topmembrane}
     where $\Phi$ is the canonical three form on the $G_2$ manifold.  
     Naively the action (\ref{topmembrane}) should reduce to (\ref{topAmodel}) 
      for wrapped configurations on $CY_3 \times S^1$ 
       \cite{Bonelli:2005ti} and this was our initial motivation for the choice 
        of (\ref{topmembrane}).

Indeed, contrary to the standard two dimensional sigma model,
the BRST quantization of the action (\ref{topmembrane}) for generic $\Sigma_3$ is a difficult task.
Actually, setting a fully covariant gauge fixing seems to couple the membrane to gravitational moduli in a 
non trivial way.
This implies two technical problems. 
The first is that ghost and antighost sectors are unbalanced and it is difficult to us 
to find natural projectors to fix it in a generic situation.
The second concerns coupling 3D topological gravity in an appropriate way.
Without doubt this is a hard problem and presently we do not have a complete understanding of it. 

However we are able to carry out the program in the case where $\Sigma_3$ is an $S^1$- bundle over a 
Riemann surface $\Sigma_g$ (Seifert manifold\footnote{Actually the notion
of Seifert manifold is more general and includes also orbifold generalizations
\cite{Beasley:2005vf}. Here
for simplicity we focus on Seifert manifolds admitting a free $U(1)$ action.}).  
Indeed the resulting theory counts associative maps, which correspond to membrane 
instantons on $G_2$ manifolds.

Hopefully the present work puts on
firmer ground some results which were derived in on shell models by Harvey and Moore 
\cite{Harvey:1999as}  and by Beasley and Witten \cite{Beasley:2003fx}. 
Actually we give a clear and complete definition of quantum amplitudes via an
off-shell procedure.
Among other results, we confirm the prescription of calculating the Euler characters of the moduli space
of associative cycles given by Beasley and Witten \cite{Beasley:2003fx}.

A related issue has to do with the interpretation of our theory as the microscopic description of 
topological M-theory. The topological membrane theory that we propose in this paper has the nice 
feature of reducing to the A model on a Calabi-Yau threefold $CY_3$ once considered on seven 
manifolds\footnote{This generalizes easily to seven manifolds $S^1\times M_6$, where $M_6$ has $SU(3)$
structure.} of type $S^1\times CY_3$. 
This will be established via equivariant localization of the membrane path-integral for Seifert geometries.
We show that the 
{\it non local} membrane
equivariant observables reduce to {\it local} A model observables.
Moreover, our reduction to the A model directly reproduces Gopakumar-Vafa invariants. 

Actually, it is only the winding sector of the membrane which gets reduced to the A model.
The unwinding sector instead stays as a topological theory of membranes in $CY_3$ which should represent 
the nonperturbative completion of the A model itself.
Once evaluated on $CY_3$ whose cycles are rational homology spheres, 
this sector generates Joyce torsion invariants \cite{Joyce:1999tz}. 
These are obtained from a very simple path-integral argument.

Another proposal for a microscopic description of topological M theory has been
suggested by de Boer et al. \cite{deBoer:2005pt,deBoer:2006bp} in terms of topological strings with $G_2$ target space.
The closed string model on $CY_3\times S^1$ reduces in this case to the B model.
This suggests to interpret the conjectured S-duality in topological strings
as an electro-magnetic duality in topological M theory, which exchanges strings
and membranes in seven dimensions.

The plan of the paper is the following. 
In Section \ref{membrane} we will describe covariantly membrane instantons
and discuss a few related classical issues.
In Section \ref{gfaction} we discuss the membrane theories for Seifert three dimensional geometries
with generic $G_2$ holonomy target.
In Section \ref{explicit} 
 we define and calculate the membrane path integral reducing it to an integral over 
the moduli space of associative maps.
In Section \ref{observables} we comment about the observables and their quantum amplitudes.
In Section \ref{reductionAmodel} we study wrapped sector of the model on $S^1\times CY_3$. 
 We discuss how our calculation is related to the zero genus 
  Gopakumar-Vafa invariants.
In Section \ref{joyceinvariant} we calculate the membrane on $CY_3$ 
 and show the relation with Joyce invariants and
torsion factors.
Section \ref{summary} is left for few concluding comments.

\section{Covariant membrane instantons on $G_2$-manifolds}
\label{membrane}

In this section we give a covariant description of membrane instantons in seven dimensional manifolds
with $G_2$-structure. These are the calibrated three-cycles.
Membranes are embeddings of reference geometries in the target, that is $X:\Sigma\to M$.
Given an invariant closed 3-form $\Phi$
on a Riemannian seven manifold $M$, then the cycle $X(\Sigma)$ is calibrated \cite{calib}
if the volume element with respect to the induced metric and the pullback of $\Phi$ on the cycle coincide.

In \cite{Bonelli:2005rw} a non covariant description of membrane instantons was given as solutions of the 
equation
\beq
    \dot{X}^\mu \pm \Phi^\mu_{\,\,\,\nu\rho} \d_1 X^\nu  \d_2 X^\rho = 0,
\eeq{noncovinst}
where the membrane 
   world-volume is of the form $\Sigma_2 \times S^1$ (also
     $S^1$ can be replaced by either an interval or a real line) with $\Sigma_2$ being a Riemann
      surface. 
In \cite{Bonelli:2005rw} it has been proved that the above equation is equivalent to the calibration 
condition 
\beq
d \ {\rm vol} (\Sigma) = \mp\frac{1}{6} X^*(\Phi).
\eeq{calibration}

In this section we will give a natural covariant description of membrane 
instantons which is valid for any topology.
Let us consider the solutions to the equation
\beq
dX^\mu \pm \Phi^\mu_{\,\,\,\nu\rho} *_{\Sigma} dX^\nu \wedge dX^\rho = 0,
\eeq{elegant}
where $*_{\Sigma}$ is the Hodge dual w.r.t. a given world-volume metric on the membrane 
$h_{\alpha\beta}$.
 It is straightforward to show that
the above equations imply the calibration condition (\ref{calibration}) and
the equation for the induced world-volume metric is  
\beq
h_{\alpha\beta}=\d_\alpha 
X^\mu g_{\mu\nu} \d_\beta X^\nu\equiv \gamma_{\alpha\beta} \ \ .
\eeq{induced}
In order to show this, let us multiply (\ref{elegant}) by the world-volume Hodge
star, getting in components\footnote{We work in the conventions where $\epsilon_{012}=1$.} 
\beq
  \sqrt{h} \ \epsilon_{\alpha\beta}^{\,\,\,\,\,\,\gamma} \d_\gamma X^\mu \pm 
\Phi^\mu_{\,\,\,\nu\rho} \d_\alpha X^\nu \d_\beta X^\rho = 0 ,
\eeq{covinst}
where $h_{\alpha\beta}$ is a generic world-volume metric.
Then by multiplying (\ref{covinst}) by $g_{\sigma\mu}\d_\delta X^\sigma$
one gets
\beq
\gamma_{\alpha\beta} = \mp \frac{h_{\alpha\beta}}{\sqrt{h}} \ X^*(\Phi) ,
\eeq{acca}
 which is the relation between the auxiliary and induced metrics. For a generic non-constant map $X$ and non-degenerate 
auxiliary world-volume metric $h_{\alpha\beta}$
this relation is non-singular.

Moreover, by multiplying (\ref{covinst}) by 
$\Phi_{\mu\sigma\tau} \d_\delta X^\sigma \d_\eta X^\tau$
and using the vector cross product identity\footnote{Notice that 
\beq
\Phi^{\mu\nu\rho}\Phi_{\mu\sigma\tau}= \delta^{[\nu\rho]}_{\sigma\tau} - *\Phi^{\nu\rho}_{\sigma\tau}
\eeq{label}.}
\beq
(\d_\alpha X \times \d_\beta X)\cdot(\d_\delta X \times \d_\eta X)
= \gamma_{\alpha\delta} \gamma_{\beta\eta} -
\gamma_{\alpha\eta} \gamma_{\beta\delta}
\eeq{identi}
one gets 
\beq
\sqrt{h} \ \epsilon_{\alpha\beta}^{\,\,\,\,\,\gamma} \epsilon_{\gamma\delta\eta}
X^*(\Phi) \pm 
\left(\gamma_{\alpha\delta} \gamma_{\beta\eta} -
\gamma_{\alpha\eta} \gamma_{\beta\delta} \right)=0,
\eeq{prequaz}
which upon using 
\beq
\epsilon_{\alpha\beta}^{\,\,\,\,\,\gamma} \epsilon_{\gamma\delta\eta}
= \frac{1}{h}\left( h_{\alpha\delta} h_{\beta\eta} - h_{\alpha\eta}h_{\beta\delta} \right)
\eeq{contract}
and (\ref{acca}) gives the calibration condition (\ref{calibration}).
Then plugging back the calibration condition in (\ref{acca})
one finally gets the induced metric equation 
(\ref{induced}). 
To check that the system is not overdetermined, then one just checks 
that the sum of all the squares 
of (\ref{covinst}) does not give further conditions.

Let us now comment on the relationship between the covariant 
(\ref{elegant}) and non covariant (\ref{noncovinst})
equations. The covariant equations read in components
\ber
&& \partial_0 X^\mu \pm \Phi^\mu_{\,\,\,\nu\rho} \frac{\sqrt{h}}{2} \ \epsilon_{0}^{\,\,\, ab}
\partial_a X^\nu \partial_b X^\rho = 0 \label{time} \ \ , \\
&& \partial_a X^\mu \pm \Phi^\mu_{\,\,\,\nu\rho} \sqrt{h} \ \epsilon_a^{\,\,\,b0}
\partial_b X^\nu \partial_0 X^\rho = 0 ,
%\label{space} 
\ \ .
\eer{space}
where the $a,b=1,2$ indexes run over the spatial components.  For the rest of the paper we 
 choose sign plus in the associative map equation.
  
Assuming that the world-volume of the membrane is of the form
$\Sigma_3=S^1\times\Sigma_2$, one can partially
fix the diffeomorphism symmetry of (\ref{elegant}) by setting
\ber
h_{00} &=& {\rm det} \ \hat h_{ab} \ \ , \nonumber\\
h_{0a} &=& 0
\eer{diffeos}
Then by inserting (\ref{time}) into (\ref{space}) we get the induced metric
condition $\gamma_{\alpha\beta}= \hat h_{\alpha\beta}$.
Finally, using this condition it is easy to verify that the time
component of the covariant equation (\ref{time}) reduces precisely
to (\ref{noncovinst}).

This shows that the non-covariant equations (\ref{noncovinst}) follow from 
(\ref{elegant}) in the case $\Sigma_3=S^1\times\Sigma_2$.
We underline that for membranes with this topology 
it is possible to fix diffeomorphisms in such a way that
the induced metric \textit{decouples} and the membrane instanton equations
can be written without reference to any world-volume metric.
This is a relevant issue for the quantization of the membrane as we 
will further comment in the next section.

Finally, let us notice that a completely analogous construction can also be designed for calibrated 
four-cycles for an eight dimensional manifold with $Spin(7)$-structure.

\subsection{Local structure}
\label{local}

Let us  study the local structure of the solutions of our instanton equations.
As we have just proved, the image of associative maps
spans associative cycles. The purpose of this subsection is to show 
this from a complementary constructive point of view.

The local $G_2$ manifold geometry in the vicinity of an associative 3-cycle $M_3$
is given by the total space of the spin bundle over $M_3$ \cite{Bryant,McLean}.
Adapting the coordinate choice to the local geometry as $\{x^\mu\}=\{x^\alpha,x^A\}$ 
with $\alpha=1,2,3$ and $A = 1,2,3,4$, the
3-form defining the $G_2$ structure expands as
$$
\Phi= - \sqrt{g_3} \ \epsilon_{\alpha\beta\gamma}dx^\alpha\wedge dx^\beta \wedge
dx^\gamma +(\gamma_{\alpha})_{AB}\,\,dx^\alpha\wedge  dx^A\wedge dx^B +O(|x^A|) ,
$$
where $\gamma_\alpha$ are the gamma--matrices in three dimensions satisfying the Clifford algebra
\break
$\{\gamma_\alpha,\gamma_\beta\}=2(g_3)_{\alpha\beta}$.
The instanton equations then reduce to
\ber
&& *dX^\alpha - \epsilon^\alpha_{\ \ \beta\gamma} \sqrt{g_3} \ dX^\beta \wedge dX^\gamma = 0,
\label{id} \\
&& *dX^A + (\gamma_{\alpha})^A_{ \ B} \ dX^\alpha  \wedge dX^B = 0.
\eer{certo}
The first equation (\ref{id}) is 
identically satisfied since $X^\alpha$ span a covering of the associative cycle $\Sigma \to M_3$.
The second (\ref{certo}) is solved by maps
$X^A =0$ up to linear deformations
described by 
$$
* \nabla \delta X^A + (\gamma_{\alpha})^{A}_{ \ B} dX^\alpha \wedge \nabla\delta X^B = 0 .
$$
Solutions of the above equation correspond to zero modes of the twisted McLean Dirac operator
\cite{McLean}.

In these adapted coordinates, our model makes contact with some previous discussions
on the semiclassical quantization of membranes \cite{Harvey:1999as,Beasley:2003fx}.

\section{Gauge-fixing the Membrane}
\label{gfaction}

\subsection{Trying to gauge fix the general model}

In this Section we consider the gauge fixing for the following topological 
membrane theory
 \beq
  S_{top} = - \frac{1}{6}\int\limits_{\Sigma_3} \,\,X^*(\Phi),
 \eeq{topaction}
 where $\Phi$ is the closed three form associated to a $G_2$-structure on a seven dimensional manifold 
$M_7$. We start by discussing the covariant instanton description of the previous section as a gauge fixing.
The topological gauge symmetry of the action is 
  \beq
   \delta X^\mu = \epsilon^\mu .
  \eeq{gaugetransf}
  The corresponding BRST operator $s$ is  defined as follows  
  \beq
  sX^\mu = \Psi^\mu,\,\,\,\,\,\,\,\,\,\,
  s\Psi^\mu =0,\,\,\,\,\,\,\,\,\,\,
%  s\bar{\psi}^\mu_\alpha = b^\mu_\alpha,\,\,\,\,\,\,\,\,\,\,
%  sb^\mu =0,
  \eeq{BRST}
   where $\Psi^\mu \in X^*(TM)\otimes \Omega^0(\Sigma)$ is the ghost associated to $\epsilon^\mu$
   and has ghost number one. 
In order to fix this symmetry, we could choose the following covariant gauge function
\beq
 {\cal F}^\mu_\alpha = \d_\alpha X^\mu + \frac{\sqrt{h}}{2}
\epsilon_\alpha^{\beta\gamma} \Phi^\mu_{\,\,\,\nu\rho} \d_\alpha X^\nu \d_\beta X^\rho =0.
\eeq{gaugefunct}
We observe that the square of (\ref{gaugefunct}) gives rise to the bosonic action
\ber
S_{bos}&=&
-\frac{1}{6} \int X^*(\Phi)+\frac{1}{3}
\int d^3\sigma \sqrt{h}h^{\alpha\beta}
\frac{1}{2}{\cal F}^\mu_\alpha g_{\mu\nu} {\cal F}^\nu_\beta \nonumber\\
&& = \frac{1}{6}\int d^3\sigma \sqrt{h}
\left[
h^{\alpha\beta}\gamma_{\alpha\beta} 
+ \frac{1}{2} \left(h^{\alpha\beta}\gamma_{\alpha\beta}\right)^2
- \frac{1}{2}h^{\alpha\beta}\gamma_{\beta\gamma}h^{\gamma\delta}\gamma_{\delta\alpha}
\right] .
\eer{square}
The stationary value of $S_{bos}$ with respect to variations of $h_{\alpha\beta}$ is reached 
for $h_{\alpha\beta}=\gamma_{\alpha\beta}$
at
\beq
S_{bos}[h=\gamma]=\int d^3\sigma\sqrt{\gamma}
\eeq{mambu}
that is the Nambu-Goto action. Henceforth, (\ref{square}) could 
be regarded as a generalization for the membrane of the Polyakov action.

In order to implement the covariant gauge-fixing (\ref{gaugefunct}) one has to 
introduce a couple of antighosts and lagrangian multipliers 
$(\bar\Psi_{\alpha\ \mu}, B_{\alpha\ \mu})$ with ghost numbers $(-1,0)$ 
respectively. These fields are sections of the bundle $X^*\left(T^*M \right)\otimes \Omega^1(\Sigma)$.
The ghost and antighost sector of our theory are therefore unbalanced. This follows
from the fact that the covariant gauge fixing (\ref{gaugefunct}) we choose is actually redundant and 
extra gauge 
conditions have to be implemented in the antighost sector, 
either via the introduction of ghosts-for-ghosts or of suitable projection operators.

This is similar to what happens in the topological sigma model case, where indeed
the antighosts are constrained by using the complex structure projectors 
\cite{Witten:1988xj, Baulieu:1989rs, Birmingham:1991ty}. 
However, in the topological membrane case a further complication takes place,
since in the covariant equations for associative cycles (\ref{elegant})
the calibration conditions (\ref{calibration}) cannot be fully disentangled from the induced
metric conditions (\ref{induced}). 
This suggests that for a fully covariant
description of the membrane one could need the coupling with topological gravity.

Nonetheless, for membranes whose world-volume is a circle fibration over a Riemann
surface (Seifert manifolds) one can easily construct suitable projection
operators on the antighost sector and write a completely gauge fixed sigma model.
Since this covers a wide class of physically relevant three--dimensional manifolds, 
as for example rational homology spheres, we find it
interesting to focus on this case.

\subsection{Gauge-fixed action for Seifert membranes}

A Seifert manifold $S_{g,p}$ is a three-dimensional manifold given by the total
space of a circle bundle over a Riemann surface $\Sigma_g$ 
\beq
   \begin{array}{ccc}
 S_{g,p}& \stackrel{}{\longleftarrow} &{S^1}\\ 
{\scriptstyle \pi}\Big\downarrow &&\\
\,\,\,\Sigma_g &&
\end{array}  
\eeq{seifert}
the $U(1)$ group acts as rotations on the fibers of (\ref{seifert}).
The topology of $S_{g,p}$ is completely specified in terms
of the genus $g$ of the Riemann surface and the degree $p$
of the bundle.
The Seifert manifolds admit a globally defined non-vanishing invariant one-form
$\kappa$, which can be taken to be a connection of the principal
$U(1)$ bundle with curvature
\beq
{\rm vol}_\omega d\kappa = p \ \pi^* \omega ,
\eeq{dekappa}
where $\omega$ is a symplectic
form on $\Sigma_g$, such that $\int_{\Sigma_g} \omega = {\rm vol}_\omega$.
The dual of $\kappa$ is the fundamental vector field $k$ which generates the $U(1)$ action on $S_{g,p}$  and satisfies $i_k \kappa=1$. From these definitions it follows
\beq
\int\limits_{S_{g,p}} \kappa \ d \kappa = p .
\eeq{vol-form}
The metric on $S_{g,p}$ can be written in the form
\beq
ds^2 = \pi^* ds^2_{\Sigma_g} + \kappa\otimes\kappa .
\eeq{seif-metric}
By taking the Hodge dual of $\kappa$ w.r.t. this metric, one gets
the useful relation
\beq
 *\kappa = \pi^* \omega .
\eeq{kappa-dual}
In local coordinates one has $\kappa=dt + a$, where
$t$ is a local coordinate on the circle and $a$ is a
$U(1)$ connection on the base. Notice that the above
splitting is defined up to local gauge transformations
\beq
t\to t+\lambda \ \ \ ,  \quad\quad\ \ \ a\to a - d\lambda,
\eeq{gaugesymm}
while $\kappa$ stays invariant.

Thanks to the existence of $\kappa$, one can easily perform a projection on the antighost
sector in order to remove the redundancy of the covariant gauge-fixing condition
(\ref{elegant}). To this purpose, it is enough to project the one-forms
$(\bar\Psi_{\mu\alpha},B_{\mu\alpha})$ along $\kappa$, or equivalently, to rewrite the gauge fixing
condition as
\beq
{\cal F}^\mu =  (*\kappa) \wedge  dX^\mu + \Phi^\mu_{\,\,\,\,\nu\rho} \ \kappa\wedge  dX^\nu \wedge dX^\rho \ \ .
\eeq{kappa-gf}
In (\ref{kappa-gf}) we moved the Hodge star operator on the first term
to simplify the subsequent computations.
 From now on we do not write explicitly $\wedge$-product and it is assumed to be present in
  appropriate places. 
 To the gauge-fixing condition (\ref{kappa-gf}) we associate a couple
of antighosts and Lagrangian multipliers $(\bar\Psi_\mu ,B_\mu )$ which are 
zero-forms on the world-volume, such that the ghost and antighost sector
are now perfectly balanced.
The BRST symmetry reads
\beq
  sX^\mu = \Psi^\mu,\,\,\,\,\,\,\,\,\,\,
  s\Psi^\mu =0,\,\,\,\,\,\,\,\,\,\,
  s\bar{\Psi}_\mu = B_\mu,\,\,\,\,\,\,\,\,\,\,
  sB_\mu =0,
  \eeq{BRST1}
and the gauge-fixing action in delta gauge
\beq
S_{g.f.} = \int\limits_{S_{p,g}} s\, \left[ \bar\Psi_\mu 
( *\kappa \ dX^\mu + \Phi^\mu_{\,\,\,\,\nu\rho} \ \kappa \ dX^\nu dX^\rho )\right] .
\eeq{kappa-act}
By evaluating explicitly this action on manifolds with $G_2$ holonomy, one gets
\beq
S_{g.f.} = \int\limits_{S_{p,g}} B_\mu {\cal F}^\mu - \bar\Psi_\mu
   \left( 
    *\kappa \nabla \Psi^\mu + 2 \Phi^\mu_{\,\,\,\,\nu\rho} \ \kappa \ dX^\nu    
     \nabla\Psi^\rho + \Psi^\sigma \Gamma_{\sigma\,\,\,\lambda}^{\,\,\,\mu}{\cal F}^\lambda
   \right) .
\eeq{nome}
The last term in (\ref{nome}) seems to spoil general covariance on the target space.
Actually as it is well explained in \cite{Birmingham:1991ty}, the covariant BRST
transformations can be obtained by a simple field redefinition, which is suggested
already by (\ref{nome}) to be 
$B_\mu' = B_\mu - \Gamma^\lambda_{\ \ \sigma\mu} \bar\Psi_\lambda \Psi^\sigma$.

By integrating by parts and using $d*\kappa = 0$, which follows from (\ref{kappa-dual}),
we can symmetrize the fermionic kinetic operator of (\ref{nome}) as
\beq
S_{ferm}^{kin} = -\frac{1}{2} \int\limits_{S_{p,g}} \left(\bar\Psi_\mu D \Psi^\mu +
\Psi^\mu D^\dagger \bar\Psi_\mu\right),
\eeq{kin-ferm} 
where
\ber
D &\equiv& *\kappa \nabla + 2 \Phi^\mu_{\,\,\,\,\nu\rho} \ \kappa \ dX^\nu \nabla
\label{D-ferm},\\
D^\dagger &\equiv& *\kappa \nabla + 2 \Phi^\mu_{\,\,\,\,\nu\rho} \ \kappa \ dX^\nu \nabla
%\nonumber \\
%  &&        
 + 2 \Phi^\mu_{\,\,\,\,\nu\rho} \ d\kappa \ dX^\nu
           - 2 \Phi^\mu_{\,\,\,\,\nu\rho} \ \kappa \ \nabla dX^\nu .
\eer{D-dagger}
Notice that $D=\delta {\cal F}/\delta X |_{{\cal F}=0}$
so that the $D$ zero-modes span the tangent directions to the moduli space of associative 
cycles.
%More compactly we will write
%\beq
%S_{ferm}^{kin}= -\frac{1}{2} \int\limits_{S_{p,g}} \Xi^\dagger{\cal D}\Xi
%\quad \ \ 
%{\rm with} \quad \ \ \Xi \equiv\left(\Psi^\mu,\bar\Psi_\mu\right) \ ,
%{\cal D}\equiv \left(\matrix{0 & D^\dagger \cr D & 0}\right) \ .
%\eeq{eccola}
Moreover in the presence of fermionic zeromodes the $\delta$-gauge is clearly not complete.
In the next sections, while evaluating the membrane path integral, we will improve it to the general case.

\subsection{Hamiltonian formalism on Seifert geometries and relation \break with the Nambu-Goto membrane}

Because of the presence of the fundamental vector field $k$, Seifert manifolds
are naturally suited for Hamiltonian quantization. In fact it is easy to
check that the first term of (\ref{kappa-gf}) can be rewritten as
\beq
*\kappa \ dX^\mu = \sqrt{h_S} \ \ {\cal L}_k X^\mu \ d^3\sigma ,
\eeq{Lie}
where $h_S$ is the determinant of the Seifert metric (\ref{seif-metric}).
Hence the first term of the projected gauge-fixing conditions (\ref{kappa-gf})
is the Lie derivative of the embedding map $X$ w.r.t. the fundamental
vector field $k$, which is the natural generalization
of the "time" derivative for Seifert manifolds. 
In particular, for trivial circle bundles it is immediate
to recognize that (\ref{kappa-gf}) reduces to the non-covariant gauge-fixing
(\ref{noncovinst}). In this subsection we discuss the Hamiltonian formalism
for the Nambu-Goto membrane and show how this relates with the topological
model discussed so far. 

We then start from the Nambu-Goto action (\ref{mambu}) and compute
the momentum associated with ${\cal L}_k X$
\beq
p_\mu = \frac{\delta S_{NG}}{\delta ({\cal L}_k X^\mu)} = \sqrt{\gamma} \ \kappa_\alpha
\gamma^{\alpha\beta} g_{\mu\nu} \partial_\beta X^\nu .
\eeq{momentum}

This satisfies the following primary constraints
\beq
p^2-\gamma\kappa_\alpha\gamma^{\alpha\beta}\kappa_\beta=0
\quad{\rm and}\quad
{\cal L}_{v_{(i)}}X^\mu p_\mu=0,\quad i=1,2 ,
\eeq{hamcon}
where the two vectors $v_{(i)}$ are linear independent and orthogonal to $\kappa$, that is
$i_{v_{(i)}}\kappa=0$.

The phase space action of the system is 
\beq
S_{ham}=\int\limits_{S_{g,p}} d^3\sigma
\left[
p_\mu {\cal L}_k X^\mu 
- N\left(
p^2-\gamma\kappa_\alpha\gamma^{\alpha\beta}\kappa_\beta
\right)
-\sum_{i=1,2}N_{(i)}{\cal L}_{v_{(i)}}X^\mu p_\mu
\right] .
\eeq{ham-act}

To get a gauge-fixed action we can fix the Lagrange multipliers to
$N_{(i)}=0$ and $N=\frac{1}{\omega_{12}}$ in terms of 
the volume density $\omega_{12}$ on the base Riemann surface. 
As in the usual case the action (\ref{ham-act})
is quadratic in the momenta. We can then integrate them out
by fixing them to their classical values. This leads to the Lagrangian action
\beq
S_{Lag}=\frac{1}{4}
\int\limits_{S_{g,p}} \kappa\omega 
\left[
\left({\cal L}_k X\right)^2+\frac{4}{(\omega_{12})^2}
\gamma\left(\kappa_\alpha\gamma^{\alpha\beta}\kappa_\beta\right)
\right] ,
\eeq{lag-act}
where $\omega=\omega_{12}d\sigma^1d\sigma^2$.

Let us now show that in this gauge, up to an additive topological term, 
(\ref{lag-act}) is the square of our instanton equation.
Actually, writing 
\be
f^\mu={\cal L}_k X^\mu+\Phi^\mu_{\,\,\,\nu\rho}\partial_\alpha X^\nu\partial_\beta X^\rho
\epsilon^{\alpha\beta\gamma}\kappa_\gamma ,
\eeq{blablabla}
(where $\epsilon^{123}=1/\omega_{12}$)
we get 
$$
f^2=\left({\cal L}_k X\right)^2+\frac{4}{(\omega_{12})^2}
\gamma\left(\kappa_\alpha\gamma^{\alpha\beta}\kappa_\beta\right)
+\frac{4}{\omega_{12}}X^*(\Phi)_{123}
$$
and comparing with (\ref{lag-act}), we get 
\beq
S_{Lag}=\frac{1}{4}\int_S \kappa\omega f^2 -\int_S X^*(\Phi),
\eeq{compare}
which shows that associative maps satisfying $f^\mu=0$ are Nambu-Goto instanton membranes.

\subsection{Coupling to gravity}

The topological model for the membrane we presented in the previous sections plays the analogous r\^ole for
topological M theory as the topological sigma model for topological
strings.
The formulation of a complete theory involves the sum over the world-volume metrics
$h_{\alpha\beta}$. To this end, one could introduce a gravitational
BRST multiplet containing the BRST partner of the metric
\beq 
s h_{\alpha\beta} = \chi_{\alpha\beta} \ \ , \quad\quad \ \  
s\chi_{\alpha\beta}=0 \ \ ,
\eeq{grave}
and a suitable gauge--fixing condition. As we already noted in Section \ref{membrane}, 
the covariant instanton equation (\ref{elegant})
implies also the condition on the world-volume metric (\ref{induced})
which could be used as a gravitational gauge--fixing. Indeed we saw
that (\ref{elegant}) is a redundant gauge condition with respect to
the symmetry (\ref{gaugetransf}) and this leads to an unbalanced
antighost sector. Unfortunately this problem is not completely
fixed by the gravitational topological symmetry since the antighost
sector count twenty-one components while the BRST ghost are only 
seven from (\ref{BRST}) plus six from (\ref{grave}).

Once again the situation is simpler for Seifert
manifolds. In this case the very form of the
metric (\ref{seif-metric}) indicates that 
the coupling with topological gravity would contain the integration
over the moduli space given by the total space of   
the fibration 
$$
 \begin{array}{ccc}
 \widehat{\cal M}_g& \stackrel{}{\longleftarrow} &{\bf T^g}\\ 
{\scriptstyle }\Big\downarrow &&\\
{\cal M}_g &&
\end{array}  
$$
where ${\cal M}_g$ is the moduli space of the Riemann surface $\Sigma_g$ and
${\bf T^g}$ the moduli space of $U(1)$ flat connections on $\Sigma_g$.
 The above diagram hints about a relation with the Gopakumar-Vafa invariants.
  Presently we do not know how to treat properly the gravitational moduli for Seifert geometries.
   However we will be able to give a definite prescription for genus zero, see Section \ref{reductionAmodel}.

\section{Explicit evaluation of the membrane path-integral}
\label{explicit}

In this section we use the topological symmetry to evaluate the membrane path integral
explicitly.

Let us start by modifying the gauge fixed action (\ref{kappa-act}) as follows
\beq
S_{top}+ s \int\limits_{S_{g,p}} \left (\frac{1}{\beta}\bar\Psi{\cal F} + \bar\Psi g^{-1} B\right ),
\eeq{semi1}
where $\beta$ is a c-number parameter that will be used to drive the exact semiclassical expansion 
while the second term in the gauge fermion resolves the fermion zero-mode ambiguity. 
The metric appearing in the last term is the $G_2$ holonomy one with respect to which 
the three-form $\Phi$ is covariantly constant.

Let us now expand our fields as follows.
Denote the full set of solutions of ${\cal F}^\mu=0$ by $X_0$
and expand 
\beq
X=X_0+\beta x ,\quad
\Psi=\Psi_0+ \beta^{1/2}\psi ,\quad
B=B_0 + b, \quad
\bar\Psi= \bar\Psi_0 + \beta^{1/2}\bar\psi,
\eeq{semi2}
where the above decompositions are defined starting from the fermionic kinetic operator ${\cal D}$ at the generic 
associative map $X_0$, namely ${\cal D}_0={\cal D}|_{X=X_0}$. 
Then, $x$ and $b$ span the space of non zero modes of ${\cal D}_0$ and
$\psi$ and $\bar\psi$ span a copy with grassmanian statistics of the same space.
Moreover, 
$\Psi_0$ are zero modes of $D_0$,
%and $B_0$ 
$\bar\Psi_0$ zero modes of $D_0^\dagger$,
while $B_0 $ is a general solution of 
\beq
D_0^\dagger B_0 = \bar\Psi_0 \left(
\frac{\delta D}{\delta X}\right)|_{X_0} \Psi_0 \ .
\eeq{classica}
As such it can be written as 
$B_0= \tilde B_0 + \hat B_0$ were $\hat B_0$ is a particular solution
of (\ref{classica}) and $\tilde B_0$ is a general solution of the associated homogeneous
equation $D_0^\dagger \tilde B_0 = 0$.

The path integral measure is then
$$
{\cal D}[X,\Psi,\bar\Psi,B]=
d[X_0,\Psi_0,\bar\Psi_0,B_0]{\cal D}[x,\psi,\bar\psi,b]
\beta^{\frac{1}{2}Tr J}
$$
where $J=\left(\matrix{1 & 0 \cr 0 & -1 \cr}\right)$ on the space spanned by $(x,b)$.
Expanding in non-zero modes of ${\cal D}_0$, namely
$D_0 x_n =\lambda_n b_n$ and $D_0^\dagger b_n = \lambda_n x_n$ (where $\lambda_n\not=0$),
we see that the $b_n$s and the $x_n$s are in one to one correspondence
and therefore $Tr J=0$.

Let us now plug the above expansion in the gauge fixed action (\ref{semi1}) so to get
\beq
S_{top}+
\int\limits_{S_{g,p}}
\left[
b D_0 x + \bar\psi D_0 \psi +B_0^2 +b^2 + R\bar\Psi_0^2\Psi_0^2
\right] + o(\beta^{1/2}) ,
\eeq{semi3}
 where $R$ stands for the curvature tensor.  
We can use then the BRST symmetry $s$ to localize our path integral, namely to evaluate it 
at $\beta\to 0$. 
In order to further simplify the evaluation of the path integral at $\beta=0$
we can get rid of the $b^2$ term in the action by the rescaling
$(x,b)\to(\mu x, \mu^{-1}b)$ and send $\mu\to\infty$. Henceforth the integral
over the fluctuations of the fields gives simply one because of exact boson-fermion
cancellation. This result is clearly expected because of the isomorphism between
the ghost and antighost sectors which makes our theory an $N_T=2$ topological
model \cite{giorgio}.

The integral over $B_0$ is Gaussian and gives two contributions:
one is a constant overall factor 
$\int d\tilde B_0 {\rm e}^{-\tilde B_0^2}$. The other gives a further
quartic term\footnote{Notice that this is an usual term arising 
in topological field theories as for example in a twisted version
of ${\cal N}=4$ SYM in four dimensions.} 
in the fermionic zero modes ${\rm e}^{-\hat B_0^2}$.  
The integral over $X_0$, namely the space of solutions of ${\cal F}^\mu=0$, 
is then an integral over the moduli space of associative maps.  The different instanton
 sectors are classified topologically by the homology class
\beq
 N = X_*(\Sigma_3) \in H_3(M, \mathbb{Z}).
\eeq{homologyclass}
Sometimes it is convenient to introduce a basis $[e_i]$ of $H_3(M, \mathbb{Z})$, where 
 $i=1,..., b_3(M)$. Thus we can expand $N= \sum\limits_{i=1}^{b_3} N_i [e_i]$ and the instanton 
  sectors are labeled by $b_3(M)$ integers $N_i$.   
Let us denote the moduli space associative maps as  ${\cal M}_N$
 for $N \in H_3(M)$.
The integral over $\Psi_0$ is therefore along $T{\cal M}_N$.
Due to the three-dimensional index theorem
$dim ( Ker \, D_0 ) = dim ( Ker \, D_0^\dagger )$,
the integration along $\bar\Psi_0$ is along a space of the same dimension.
Notice that the remaining terms in the reduced action, namely $R\bar\Psi_0^2\Psi_0^2+\hat B_0^2$, are quartic in the fermionic zero modes and therefore soak-up the fermionic zero modes.
Therefore performing the integration over zero modes $\Psi_0$ and $\bar\Psi_0$ we get the Euler class 
\beq
 e({\cal M}_N) = Pf ({\cal R}_N)~,
\eeq{eulerclass}
 where now ${\cal R}_N$ is regarded as Lie algebra valued two-form, curvature two-form on 
  ${\cal M}_N$.
    It is important to stress that we took into account the contribution from $\hat{B}_0^2$-term.
Due to the Gauss-Bonnet thorem $e({\cal M}_N)$ upon the integration over ${\cal M}_N$ (i.e., 
 the remaining integration $d[X_0]$)
 gives the Euler number $\chi({\cal M}_N)$ of ${\cal M}_N$.  Introducing the parameters 
 \beq
  t_i = \frac{1}{6} \int\limits_{e_i} \Phi
 \eeq{parametersu2829}
 with $\Phi$ being $G_2$ three form and defining $q_i= e^{-t_i}$ we can write the partition function 
 as follows
\beq
 Z= \sum\limits_{N \in H_3(M)} \chi({\cal M}_N) \,\,q^N
\eeq{partiatioandjkk}
where $q^N$ denotes $\prod\limits_{i=1}^{b_3} q_i^{N_i}$. The evaluation of $\chi({\cal M}_N)$ 
 can be a hard problem and depends very much on the structure of ${\cal M}_N$ which is not known.   
  However  in Section \ref{reductionAmodel} we will present the evaluation of $\chi({\cal M}_N)$
   for specific $N$ on $CY_3 \times S^1$.

Alternatively the result (\ref{partiatioandjkk}) can be argued through more formal 
 Mathai-Quillen formalism (see \cite{Wu:2005pr} for the review) with ${\cal F}$ being the appropriate 
  zero section.  The analysis goes along the lines presented in \cite{giorgio}.

\section{Observables and moduli spaces}
\label{observables}

We now discuss the observables of the general topological model introduced in
Section \ref{gfaction}.
Let us recall the situation as it was already 
discussed in \cite{Bonelli:2005rw}.
The construction of observables closely follows the analogous one for the 
A-model topological string. However the path integral evaluation of them 
takes a different path because of the three dimensional index theorem.

For a nontrivial element $[A] \in H^q(M)$ we 
can define the following co-cycles
\beq
{\cal C}_i^{q-i} \equiv \frac{1}{i!} A_{\mu_1...\mu_q} dX^{\mu_1} \ldots  dX^{\mu_i} 
\Psi^{\mu_{i+1}}\ldots\Psi^{\mu_{q}}
\eeq{closedforma}
for $q\geq i \geq 0$ and zero otherwise.
In (\ref{closedforma}) the upper index stands for the ghost number 
 and the lower index for the degree of the differential form on the world-volume $\Sigma$. 
 Using the transformations (\ref{BRST1}) we can derive 
 the decent equations for ${\cal C}_i^{q-i}$
\beq
d{\cal C}_{i-1}^{q-i+1} = (q-i+1) s {\cal C}_i^{q-i} .
\eeq{deceqsk}
Thus ${\cal C}_0^q$ are BRST-invariant {\it local} observables labeled by 
   the elements of the de Rham complex $H^\bullet (M)$, while 
   from ${\cal C}_i^{q-i}$ with $ i >0$ we can construct 
     BRST-invariant {\it non-local} observables as integrals
     \beq
     \int\limits_{c_i} {\cal C}_i^{q-i} \, ,
     \eeq{integracycle}
      where $c_i$ is an $i$-cycle on $\Sigma$. 

Not all observables have non-vanishing correlators in the theory.
This in general depends on the possible anomalies that the theory displays in the ghost sector.
Indeed, the gauge-fixed action (\ref{kappa-act}) has at the classical level a ghost 
number conservation law, with $\Psi$ having ghost number $1$, $\bar{\Psi}$ 
having ghost number $-1$ and $X$ having ghost number $0$. The BRST transformation  $s$ (\ref{BRST1}) 
changes the ghost number by $1$. 

Notice that all the observables, but ${\cal C}_i^0$ with $i=1,2,3$, defined in (\ref{closedforma})
have positive ghost number. Thus, in order to have non--vanishing
correlators there should be a compensating ghost number anomaly.

The linearized equations for the fermionic fluctuations around the instanton background are
\ber
&&D_0\Psi^\mu_0
%=\nabla_0 \psi^\mu + \Phi^\mu_{\,\,\,\nu\rho} \epsilon^{ab} \nabla_a \psi^\nu \d_b X^\rho 
=0,\label{zeromod22}\\
 &&D_0^\dagger \bar{\Psi}_{0~\mu} 
%= \nabla_0 \bar{\psi}^\mu - \Phi^\mu_{\,\,\,\nu\rho} \epsilon^{ab} \nabla_a \bar{\psi}^\nu \d_b X^\rho 
=0.
  \label{zeromod11}
\eer{zeromodelala}
The equation (\ref{zeromod22}) is the first order variation of the associative map 
(\ref{kappa-gf}). As such, $\Psi_0$   
can be interpreted as a section of the tangent bundle to 
the moduli space ${\cal M}$ of associative maps.  
The operator  $D^\dagger$ is the adjoint of $D$, 
and thus the ghost number anomaly is given by the index $ind(D)$.  
Since our theory lives in three dimensions 
 $ind(D)$ vanishes by index theorem. 
Thus the correlators of any observable in 
(\ref{closedforma}) with positive ghost number do not get contribution. Hence we are left
with the observables $\int_{c_i}{\cal C}^0_i$ with $i=1,2,3$.
For example, the case $i=3$ corresponds to our classical action (\ref{topaction}) and
its variations in $H^3(M)$.
As such its evaluation corresponds just to the
partition function, which computes the Euler characteristic of the moduli
space of associative maps \cite{Bonelli:2005rw}. 
The situation is very similar to supersymmetric quantum 
mechanics (for a review, see \cite{Birmingham:1991ty}). 

The above conclusions can be enforced also by direct calculation of the relevant quantum amplitudes 
via the method explained in the previous section. 
Suppose then we want to evaluate the insertion of any of the observables (\ref{integracycle}) or of 
a product of them.
Actually these are functionals of $X$ and $\Psi$ only
${\cal O}(X,\Psi)$. 
The change of variable (\ref{semi2}) $X=X_0+\beta x$ and $\Psi=\Psi_0+\beta^{1/2}\psi$
and the subsequent $\beta\to 0$ limit leaves 
${\cal O}(X_0,\Psi_0)$ and the insertion is independent on the fluctuations 
of the fields.
To calculate the quantum amplitude, therefore one stays with
\beq
\int d[X_0,\Psi_0,\bar\Psi_0]{\rm e}^{-S_{top} 
%+ \frac{i\pi}{2}\eta[D_0] 
 -\int_\Sigma (R\bar\Psi_0^2\Psi_0^2 + \hat B_0^2)} {\cal O}(X_0,\Psi_0).
\eeq{LESS}

Because of the equality among the number of $\Psi_0$s and $\bar\Psi_0$s there is 
a simple selection rule which states that the only observables which have not a priori
zero amplitude are the ones at zero ghost number where there is no dependence upon
$\Psi$ at all and therefore no extra $\Psi_0$ mode to be soaked up.

Let us note that the above conclusions are based on the smoothness 
of the moduli space of associative maps, that is on the stability of its tangent bundle. 
This condition might be violated by 
some components thus invalidating the above arguments.
Unfortunately not much is known about the geometry of these moduli spaces (for the recent discussion, see \cite{akbulut1, akbulut2})
and therefore we are not able to further discuss this issue.

\section{Membranes on $S^1$ and topological A model}
\label{reductionAmodel}

In this section we show the explicit reduction at quantum level of our topological 
membrane theory to the topological A model on appropriate geometries.
Let us consider the target geometry
to be $S^1\times X$, where $X$ is a Calabi-Yau threefold.
Therefore, we have $\Phi=dX^7 K+{\rm Re}\Omega$, where $K$ is the Kahler two-form and
$\Omega$ the holomorphic  three-form on $X$.
Then, as already discussed in \cite{Bonelli:2005rw}, membranes can have zero or non
zero winding along the target $S^1$.
The full quantum theory will receive two distinct contributions by these two sectors.
The zero-winding sector corresponds to membranes all internal to the CY threefold and will 
be the subject of next section. 
The non zero winding sector is our concern now and will be studied in the following.
In particular, we show how to recover {\it local} observables in the A model from {\it non local}
observables in an equivariant sector of the topological membrane theory.

The first simplification in calculating the membrane path integral
in the non-zero winding sector
comes from the presence of the isometric $S^1$ direction.
Once combined with rotations along the fiber of the Seifert geometry,
this allows us to compute the membrane path integral via equivariant localization.
Let us first define the equivariant BRST action 
\beq
s_\kappa X^\mu= \Psi^\mu ,
\quad
s_\kappa \Psi^\mu= u k^\alpha\partial_\alpha X^\mu - wV^\mu,
\quad
s_\kappa\bar\Psi_\mu = B_\mu,
\quad
s_\kappa B_\mu = u k^\alpha\partial_\alpha \bar\Psi_\mu 
\eeq{eqbrst}
and $s_\kappa u=s_\kappa w=0$. 
The vector $V^\mu=\delta^\mu_7$ generates the target space isometry
and $k^\alpha\partial_\alpha$ is the vector generating the Seifert $U(1)$-action.
Notice that $s_\kappa^2$ closes
on the $U(1)$ action that we will use to localize, namely
\beq
s_\kappa^2 X^\mu = u k^\alpha\partial_\alpha X^\mu - wV^\mu,
\quad
s_\kappa^2 \Psi^\mu = u k^\alpha\partial_\alpha \Psi^\mu ,
\quad
s_\kappa^2 \bar\Psi_\mu = u k^\alpha\partial_\alpha \bar\Psi_\mu ,
\quad
s_\kappa^2 B_\mu = u k^\alpha\partial_\alpha B_\mu .
\eeq{eqbrstsq}
We write the above as $s_\kappa^2={\cal L}_U$
on the field space.

Before entering the path integral evaluation, let us face the problem of 
defining equivariant observables.
As a first point, notice that the topological action (\ref{semi1}) itself is trivially 
$s_\kappa$-closed and $U$ invariant. Actually, it is not $s_\kappa$-exact.
As it is clear, equivariant observables are obtained by
calculating the cohomology of $s_\kappa$ in the space
of $U$-invariant functionals.  
Let us analyze a particular observable
\beq
{\cal O}=\oint\limits_{S^1}A_{\mu\nu\rho}dX^\mu\Psi^\nu\Psi^\rho+f_\mu dX^\mu.
\eeq{eqobs}
with $A \in H^3(M_7)$ and $f\in\Lambda^1(M_7)$.
By direct calculation we verify that 
\beq
s_\kappa {\cal O}=0\quad  {\rm iff} \quad df=2wi_V A =2wB,
\eeq{iff}
 where it is enough for $f$ to exist locally. 
Since $M_7=S^1\times X$ we have $i_V A = B \in H^2(X)$ and  
the solution to the above condition is provided by
picking $f$ to be the local potential of $B$ divided by $2w$. 
Because of the simple connectivity of the CY
this local choice can be extended to cover all the
image of the fiber $S^1$ in $X$ and therefore defines unambiguously
our observable (\ref{iff}). Indeed for any element $B$ of $H^2(X)$ we can define 
 the equivariant observable (\ref{eqobs}) specified by $A= dX^7 \wedge B$.
Notice that an analogous construction for the local observable $i_V A \Psi\Psi$ does not 
hold.

The evaluation of the membrane path integral by equivariant localization under the $U(1)$
action (\ref{eqbrstsq}) reduces the domain of integration to the U-fixed points.
These field configurations are the solutions of ${\cal L}_U[fields]=0$,
that is
\beq
X^\mu=\frac{w}{u}V^\mu (t-\sigma) + x^\mu,
\quad
\Psi^\mu=\psi^\mu,
\quad
\bar\Psi_\mu= \bar\psi_\mu,
\quad
B_\mu= b_\mu,
\eeq{fixpts}
where $t$ is a local coordinate on the $S^1$ fiber of the Seifert and
$\sigma= -i \ln e$, $e$ being a section of the $U(1)$ bundle over $\Sigma_g$.
Notice that when the $U(1)$ bundle is non trivial $\sigma$ becomes multivalued.
Henceforth, in order the map (\ref{fixpts}) to be regular, we 
require\footnote{We underline that this condition is necessary 
for a target space CY$\times S^1$ but could be relaxed for more
general geometries as for example $S^1$ fibrations over a CY.}
the Seifert to be $S^1\times \Sigma_g$ and thus we can choose $\sigma=0$.
In (\ref{fixpts}) $x^\mu$, $\psi^\mu$, $\bar\psi_\mu$ and $b_\mu$
depend only on the base $\Sigma_g$. Moreover, using the gauge symmetry
of the Seifert geometry (\ref{gaugesymm}), the field $x^7$ can be reabsorbed
by an appropriate choice of 
the local coordinate $t$. Correspondingly, due to the BRST invariance, also the
fermionic partner $\psi^7$ has to be set to zero. 

While reducing to the fixed locus, the induced measure 
is obtained by the product of the 
weights of the $U(1)$ action (\ref{eqbrstsq}) on the tangent 
space\footnote{Actually, for non isolated fixed points, one should \cite{hkdo} consider also the curvature of the 
normal bundle ${\cal R}_{\cal N}$. 
However, due to the mutual independence of the Fourier modes along the fiber, in our case
${\cal R}_{\cal N}=0$.}.
%
%\bibitem{hkdo}
%N.~Berline, E.~Getzler and M.~Vergne, ``Heat Kernels and Dirac Operators'', Springer, Berlin, 1996.
%
This is easily calculated to be
\beq
\left[{\rm det}^{1/2}(u {\cal L}_\kappa )\right]^{[1+1-1-1]}=1,
\eeq{jacU}
where the exponents ($+1$ for the fermions and $-1$ for the bosons) 
sum up to zero just for usual fermion/boson cancellation.

Notice that the way we are counting the field modes in the measure has changed 
w.r.t.  Sections \ref{explicit} and \ref{observables}. In fact there the measure was defined by 
diagonalizing the three-dimensional fermionic kinetic operator
in (\ref{kin-ferm}). Instead in the equivariant approach we 
diagonalize the operator ${\cal L}_U\otimes \nabla$, where $\nabla$
is the fermionic kinetic operator for the A-model on the base.
Since in general the two operators do not commute, 
these correspond to different regularization schemes of 
the path integral. In particular the selection rules discussed
in Sections \ref{explicit} and \ref{observables} have to be changed since now one 
has to take into account
the index theorem for the reduced fermionic kinetic operator $\nabla$.

Let us now proceed further by analyzing the gauge fixed action.
As a preliminary step, let us notice that 
$s_\kappa \int_\Sigma\bar\Psi_\mu{\cal F}^\mu=
s \int_\Sigma\bar\Psi_\mu{\cal F}^\mu$ 
and therefore the gauge fixed 
action stays unchanged.
The gauge fixing conditions (\ref{kappa-gf})
under localization on (\ref{fixpts}), reduce to
\ber
\label{extrastuff}{\cal F}^7|_{f.l.}  &=& \kappa f^7 = \kappa\left[\frac{w}{u}\omega + x^*(K)\right] ,\\
{\cal F}^m|_{f.l.}  &=& \kappa f^m =\kappa\left[
{\rm Re}\Omega^m_{np}dx^ndx^p\right]
\eer{nome2}
and the reduced gauge fixed action reads 
\beq
s_\kappa \int\limits_{\Sigma_g}\bar\Psi_\mu {\cal F}^\mu |_{f.l.}=
\int\limits_{\Sigma_g}
b_7 f^7 + b_m f^m
+
2 \bar\psi_7
 \ K_{mn} dx^m\nabla\psi^n 
%\\
%&&
+\bar\psi_m  {\rm Re}\Omega^m_{np}dx^n\nabla\psi^p ,
% \nonumber
\eeq{nomissimo}
where we integrated along the fiber with $\oint dt = 1$.

Let us first integrate over the $7$-sector. 
The constraint (\ref{extrastuff}) gives rise to a delta-function which 
 fixes the volume form
on the base in terms of the 
degree of the
reduced map and of the non-vanishing $S^1$ winding.
Actually the A-model on $\Sigma_g$ depends only
on conformal classes and therefore has to be independent
on the actual choice of $\omega$, which can be considered
as a gravitational modulus of the Seifert geometry.
 Indeed this situation is similar to the discussion of torus partition function 
  in A-model \cite{Bershadsky:1993ta}.
Henceforth, in order to recover the topological sigma A model   
we have to integrate on $\omega$ and its BRST partner.
This integration presents a subtlety related to degenerate
volume forms lying on the boundary $\omega=0$.
We assume the existence of a suitable BRST invariant
regularization of the $\omega$ quantum measure
avoiding this point.
 
The resulting gauge-fixed action 
reads finally
\beq
S_{g.f.}= \int\limits_{\Sigma_g}
b_m \left(
{\rm Re}\Omega^m_{np}dx^ndx^p\right)
+\bar\psi_m
\left(
{\rm Re}\Omega^m_{np}dx^n\nabla\psi^p
\right) \ \ .\nonumber
\eeq{saluti}
Let us now show how this reduces to the usual A model path integral.
First of all the bosonic constraint in (\ref{saluti}) can be 
written in complex coordinates as
\beq
\Omega^{\bar I}_{JK} \partial_z x^J \partial_{\bar z} x^K = 0
\eeq{baci}
and its complex conjugate. Clearly, since $\Omega$ is non-degenerate,
(\ref{baci}) is equivalent to the (anti-)holomorphicity of the map
$x$.
The resulting path integral is
\ber
&&\int {\cal D}[x,\psi,\bar\psi] \delta
\left( {\rm Re}\Omega dx dx \right) 
{\rm e}^{-\int_{\Sigma_g} \bar\psi {\rm Re} \Omega dx\nabla\psi}
\label{vabe}\\
&&=  \int {\cal D}[x,\psi,\bar\psi] 
{\rm det}^{-1}\left(  {\rm Re}\Omega dx \right)
\delta
\left( dx - J \ *dx \right) 
{\rm e}^{-\int_{\Sigma_g} \bar\psi {\rm Re} \Omega dx\nabla\psi}
 + [{\rm anti-holom.}] , \nonumber
\eer
where $ dx - J \ *dx=0$ is the usual holomorphicity condition
and the second factor takes into account the equal contribution
of anti-holomorphic maps.
To make a closer contact with the usual A model, we can
redefine the antighost $\bar\psi$ as
\beq
\rho_z^{\bar I}=\frac{1}{2}\bar\psi^J\Omega^{\bar I}_{JK}\partial_z x^K
\quad \ 
\rho_{\bar z}^I=\frac{1}{2}\bar\psi^{\bar J}\bar\Omega^I_{\bar J \bar K}\partial_{\bar z}x^{\bar K} \ .
\eeq{pippo}
This change of variables is singular on constant maps.
Actually, as we can see from (\ref{extrastuff}) these maps correspond 
to $\omega=0$ and thus are avoided in our regularized path integral. 
The Jacobian associated to (\ref{pippo})
is compensated by the determinant factor in (\ref{vabe})
leaving us with
\beq
 \int {\cal D}[x,\psi,\rho] 
\delta
\left( dx - J \ *dx \right) 
{\rm e}^{-\int_{\Sigma_g} \rho \nabla\psi}
+ [{\rm anti-holom.}]
\eeq{basta}
which is the A model path integral in $\delta$-gauge.
Indeed on holomorphic maps $\rho$ satisfies exactly the self-duality condition
\beq
\rho + J \ *\rho = 0
\eeq{self}
which is the usual projection on the antighosts of the A model
\cite{Witten:1988xj}. 

Let us now consider the evaluation of the equivariant observables
(\ref{eqobs}). On the fixed locus (\ref{fixpts}) this reduces
to
\beq
{\cal O}|_{f.l.} = \oint\limits_{S^1} \frac{w}{u} dt\,\, B_{mn}\psi^m\psi^n
= \frac{w}{u} B_{mn}\psi^m\psi^n ,
\eeq{ofl}
 where $B$ is defined below the equation (\ref{iff}).
So, the {\it non local observable of the membrane theory} (\ref{eqobs})
{\it reduces to the local observable of the topological A model}
of the standard type.
The topological action (\ref{topaction}) reduces consistently
to 
\beq
S_{top}|_{f.l.} = -\frac{1}{6} \int\limits_{S^1\times \Sigma_g} K_{mn} \frac{w}{u} 
dt dx^m dx^n = -\frac{1}{6}  \frac{w}{u}\int\limits_{\Sigma_g} x^*(K).   
\eeq{topiction}

To summarize, the equivariant localization of 
our membrane theory on $CY_3 \times S^1$ reproduces the topological A sigma-model.
In particular, due to the absence of gravitational moduli,
if $\Sigma_g = {\bf P}^1$ 
we can reduce  to A-model calculation on $CY_3$. 
However, compared with the standard A-model, the contribution 
    of the constant maps is missing here. 
     Indeed in such a case the constraint (\ref{extrastuff}) 
      is singular. This should come as no surprise since the map
       which has non-zero winding along $S^1$ and is constant on the base does not satisfy 
        the associative map 
         condition.

\subsection{Relation to Gromov-Witten and Gopakumar-Vafa invariants}

In this subsection we would like to summarize our equivariant calculation for the zero genus 
  and relate it  to the general expression (\ref{partiatioandjkk}). We comment also on the relation
   to the Gromov-Witten and Gopakumar-Vafa invariants \cite{Gopakumar:1998jq}.

Let us first of all remind some basic facts about the 
zero-genus A-model calculations which will be useful in the following discussion.
Consider the case when $X$ is an ideal Calabi-Yau threefold.  Following the standard
 arguments the three point function at zero genus is given by
 \beq
  \langle O_i O_j O_k \rangle = \int\limits_{X} l_i \wedge l_j \wedge l_k + \sum\limits_{C\subset X} \sum\limits_{m=1}^{\infty} n_i n_j n_k \,\,e^{-m \int\limits_{C} K},
 \eeq{threepointfunction} 
  where $l_i$ is the basis in $H^2(X)$ which label the observables $O_i$ and $n_i = \int\limits_{C} l_i$.
   In (\ref{threepointfunction}) we have to sum up over all rational curves $C$ and all possible degrees
    $m$.  Alternatively the second term in (\ref{threepointfunction}) can be rewritten as 
    \beq
     \sum\limits_{0\neq \beta \in H_2(X, \mathbb{Z})}\,\,\,\, \sum\limits_{m=1}^{\infty} n^0_\beta \,\,n_i n_j n_k \,\,   q^{m\beta} ,
      \eeq{ryeyui29899}
 where $n^0_\beta$ enumerates rational curves in $X$ of class $\beta$. As usual $q^{m\beta}$ is defined
  as follows
  \beq
   q^{m \beta} = e^{-m \beta \cdot \vec{t}},\,\,\,\,\,\,\,\,\,\,\,\int\limits_{\beta} K = \beta \cdot \vec{t}.
  \eeq{gopa}
   The free energy $F_0(q)$ is related through third derivative to $\langle O_i O_j O_k \rangle$ 
   $$ \langle O_i O_j O_k \rangle  = \frac{\d^3}{\d t_i \d t_j \d t_k} F_0(q)$$
    and thus we obtain
    \beq
     F_0 (q) = \sum\limits_{0\neq \beta \in H_2(X, \mathbb{Z})}\,\,\,\, \sum\limits_{m=1}^{\infty} n^0_\beta \frac{1}{m^3} q^{m\beta}, 
    \eeq{GVform}
     where we ignore the classical piece in (\ref{threepointfunction}). $n_\beta^0$ are integer numbers
      which are called zero genus Gopakumar-Vafa invariants.     
      Alternatively we can rewrite 
      (\ref{GVform}) as
      \beq
        F_0 (q) = \sum\limits_{0\neq \beta \in H_2(X, \mathbb{Z})}\,\,\,\,  N_\beta^0 q^\beta, 
      \eeq{GWform}
 where $N_\beta^0$ are zero genus Gromov-Witten invariants. 
 
%  So far it was just a reminder of the well-known facts about zero genus A-model. 
Let us now turn to 
   our equivariant calculation from the previous Section 6. We calculate the three point function of 
    non-local observables (\ref{eqobs}) which are labeled by the basis in $H^2(X)$. 
     Now $m$ is the winding number of the membrane\footnote{We fix the base orientation in (\ref{extrastuff}) in such way
     that $m$ takes only positive values. Switching to negative values corresponds to the consideration 
      of anti-instantons, i.e. the minus sign in our associative map condition (\ref{elegant}).},
  $m= \frac{w}{u}$ in (\ref{fixpts}) of the previous Section. 
  The equivariant calculation produces the following 
  contribution to the three point function
  \beq
 \sum\limits_{0\neq \beta \in H_2(X)} \sum\limits_{m=1}^\infty \,\,\,\,\,m^3 \,\,\left (    n^0_\beta \,\,n_i n_j n_k   \right ) \,\,q^{m\beta},
  \eeq{dofooow889}
where  the factor $m^3$ comes from the reduction of the observables (\ref{ofl}). Integrating these three point 
   correlators to a free energy of wrapped membranes produces the following result 
\beq
   \sum\limits_{0\neq \beta \in H_2(X)} \sum\limits_{m=1}^\infty \,\,\,\,\,n^0_\beta  \,\,q^{m\beta},
\eeq{corllslslsll}
 which is just counting the wrapped associative maps and since $X$ is ideal Calabi-Yau, all these
  maps are isolated.  It is natural to compare (\ref{corllslslsll}) with the general formula (\ref{partiatioandjkk}) from where we see that 
   the Euler number in the case of isolated maps is just the number of maps. 

We hope that above reasoning can be generalized to the case when $X$ is not an ideal CY. We believe 
 that the formula (\ref{corllslslsll}) should be still valid in this case.

\section{Membranes on CY's, homology spheres and Joyce invariants}
\label{joyceinvariant}

We still have to threat the zero winding sector of the membrane.
This corresponds to expand the path integral around $X^7=0$ (or any constant point on $S^1$).
Plugging this choice in the associative maps equation, one stays with the analog equation for 
membranes spanning special lagrangian submanifolds in the CY-threefolds
\beq
*\kappa g_{ab}X^*(dx^b)+\kappa X^*(i_{\partial_a}{\rm Re}\Omega)=0
\eeq{CYinstanton}
and the condition $X^*(\omega)=0$. Actually, the latter is implied by the first since it is just the 
condition of 
the cycle being lagrangian.
We can use a suitable shift in the gauge fermion 
to lift completely the $X^7,\psi^7$ sector.

As it was explained already in \cite{Bonelli:2005rw}, one can 
formulate also a membrane theory on CY threefolds. This theory localizes on the special 
 Lagrangian submanifolds of $CY_3$, see \cite{Bonelli:2005rw} for further details. 

The corresponding instanton equation for Seifert manifolds is given by (\ref{CYinstanton}).
The partition function can be obtained following the same steps we outlined 
in the previous sections for $G_2$ manifolds.
In particular, one can calculate it in the case in which
the associative cycle is a rational homology sphere.
This corresponds to a Seifert geometry built on a Riemann sphere.
Then we have, for $p\geq0$, $S_{0,p+1}=S^3/{\bf Z}_p$ which generalizes the Hopf bundle 
$S_{0,1}=S^3$.

Let us consider the case in which $\Sigma=S^3$ and the CY to be such that
every SLag is a rational homology sphere $S_{0,p_i+1}=S^3/{\bf Z}_{p_i}$
or just restrict our attention to such a geometrical sector as in
\cite{Joyce:1999tz}.

In this case, the tangent bundle at each cycle is trivial and therefore the cycles are isolated.
Calculating the partition function one gets simply the 
Euler number of each pointlike component of the moduli space, which is $1$.
Actually, in summing up to isomorphisms one has to be careful
not to get an under-counting of each component.
In fact, two maps which differ by a ${\bf Z}_{p_i}$ transformation map to the same
cycle, but are counted as distinct. 
Actually, in the orbifolded angular direction $\theta$, the maps 
$$
\theta\to n\theta+2\pi\frac{q}{p_i},\quad q=0,\ldots,p_i-1
$$
identify the same target cycle.
Therefore for any such a cycle we get a multiplicity of $|{\bf Z}_{p_i}|=p_i=
|H^1\left(S_{0,p_i+1},{\bf Z}\right)|$.

Henceforth the partition function on the CY manifold $X$ is given by
\beq
Z_{membrane}^{X}=\prod_{i=1,\ldots,h_3(X)}
|H^1\left(S_{0,p_i+1},{\bf Z}\right)|
\sum_{n_i\in{\bf N}}e^{-t_in_i}c(n_i) ,
\eeq{joyce}
where $\{t_i\}$ span the real part of the complex moduli of $X$,
$n_i$ is the degree of the map at the $i$-th cycle
and 
$c(n_i)$ is a function of such a degree that we do not calculate.
In fact, these coefficients would arise by calculating the integrals over the 
moduli space of multiple covering associative maps.
Unfortunately very little is known about such spaces.

The torsion prefactor $|H^1\left(S_{0,p_i+1},{\bf Z}\right)|$ 
is in agreement with a conjecture by D.Joyce
\cite{Joyce:1999tz}. 
In this paper, he also conjectures a prescription to account for 
multiple coverings with $c(n_i)=\frac{1}{n_i}$.

Actually, the same reasoning can be applied to membranes in $G_2$ manifolds 
with topology $S^3/{\bf Z}_{p}\times {\bf R}^4$ and produces the 
same torsion prefactor in agreement with \cite{Harvey:1999as}, 
although in such a case the multiple covering coefficient is expected to be 
$c(n)=1/n^2$ \cite{Ooguri:1999bv,Brandhuber:2001yi}.

\section{Summary and open problems}
\label{summary}

We presented a full quantum description of topological
membranes with Seifert geometry mapping into a seven manifold
 of $G_2$ holonomy. The theory is localized on associative maps and 
  the partition function calculates the Euler number of the moduli space of these maps.
   Thus the contribution of the constant maps gives zero partition function since the Euler
    number of a seven manifold is always zero. Considering the manifold $CY_3 \times S^1$
    the topological membrane has two sectors, wrapping membranes and unwrapping membranes.
     Using equivariant localization we have shown that the wrapped sector reduces to 
      the A-model calculation, except the constant map contribution.
      In particular for genus zero we show that our model computes the Gopakumar-Vafa 
invariants.  The unwrapped sector corresponds
       to the theory localized on special Lagrangian submanifolds inside $CY_3$.  
        In Section \ref{joyceinvariant} we studied in detail this sector for the case of rational 
         homology spheres getting the Joyce invariants. 
         Actually this result can be interpreted as a counting of distinct 
         D-branes in A-model, namely
         flat $U(1)$ bundles over SLags. 
This confirms that the topological membrane describes both the perturbative and non-perturbative
sector of topological strings. Concerning its second quantized description, the absence
of a non-trivial contribution by the constant maps prevents us from recovering
an Hitchin functional.  
Notice in fact that our microscopic description applies 
to a strongly coupled regime with respect to the topological $G_2$ string
proposal by de Boer et al. \cite{deBoer:2005pt}, 
 whose restriction to the constant maps gives the Hitchin functional. 
Thus indeed in our calculations this functional has
not to arise at all. However we believe that the proper interpretation  
as an effective action for topological M-theory \cite{Dijkgraaf:2004te} 
should arise in a second quantized membrane theory. 

Several open problems remain to be analyzed for a deeper
understanding of the membrane dynamics.
It would be useful to develop a viable fully covariant
gauge fixed model. As we discussed in Sections \ref{membrane} and \ref{gfaction}  this is
a relevant issue for a wider comprehension of the 
coupling to 3D gravity. 
In this direction also the relation of our topological model 
with usual super-membrane formulations via a suitable twist
should be investigated. Actually we have shown that when
written in adapted local coordinates around the associative  
cycles, our model reproduces the super-membrane theories
developed by Harvey and Moore \cite{Harvey:1999as} 
 and Beasley and Witten \cite{Beasley:2003fx} in order
to compute the membrane contributions to the superpotential
in $G_2$ M-theory compactifications.
To proceed further in these computations
within our framework, one should be able to take into
account multi-covering maps. 

Another interesting aspect to investigate concerns
the geometry of the moduli space of associative cycles
and its relation with the membrane quantum amplitudes.

A natural generalization of the present approach to the 
topological three-brane on Spin(7) geometries is to choose the
four manifold on which the three-brane is modeled to be an $S^1$
fibration over a three manifold. The program we outlined in this paper can be 
repeated in this case too. Moreover, the index theorem does not imply the 
vanishing of the ghost number anomaly and the situation is much similar to the case of the 
topological A model in two dimensions. This would be a natural starting point for the 
definition 
of some higher dimensional analogues of the Gromov-Witten invariants for complex surfaces.
This theory may serve as a microscopic description of topological F-theory \cite{Anguelova:2004wy}.

\label{end}

\noindent{\bf Acknowledgments}:  We are grateful to Sergei Gukov, Ulf Lindstr\"om,  Andrei Losev and  
George Thompson for discussions. In particular we thank George Thompson 
for reading and commenting the manuscript. 
 The research of G.B. is supported by 
 the European Commission RTN Program MRTN-CT-2004-005104 and by MIUR.
  G.B. thanks the Yukawa Institute, Kyoto where part of this work was carried out supported by the programme "Fundamental Interactions and the Early Universe". 
   M.Z. thanks SISSA (Trieste) and Simons workshop (Stony Brook)
   where part of this work was carried out. 
 The research of M.Z. was supported by  VR-grant 621-2004-3177.

\end{document}